\documentclass[11pt,fleqn]{article} 
\usepackage{graphicx}
\usepackage{textcomp} 
\usepackage{epstopdf}
\usepackage{latexsym}
\usepackage{amsmath}
\usepackage{amssymb}
\usepackage{bm}
\usepackage[mathcal]{euscript}
\usepackage{slashed}
\usepackage{tikz}

\usepackage[top=30mm, bottom=30mm, left=25mm, right=25mm]{geometry}

\numberwithin{equation}{section}

\usepackage{marvosym}

\usepackage{indentfirst}

\newcommand\blfootnote[1]{
  \begingroup
  \renewcommand\thefootnote{}\footnote{#1}
  \addtocounter{footnote}{-1}
  \endgroup
}

\usepackage{hyperref}
\hypersetup{
        unicode,	
        colorlinks,
        citecolor=blue,
        linkcolor=red, 
        urlcolor=red,
        bookmarksopen=true,
        bookmarksopenlevel=\maxdimen,
      }

\def\gl#1#2{\ifmmode \mathrm{GL}(#1; {\bf #2}) \else $\mathrm{GL}(#1; {\bf #2})$\fi}
\def\sl#1#2{\ifmmode \mathrm{SL}(#1; {\bf #2}) \else $\mathrm{SL}(#1; {\bf #2})$\fi}
\def\so#1{\ifmmode \mathrm{SO}({#1}) \else $\mathrm{SO}(#1)$\fi}

\def\sp#1#2{\ifmmode \mathrm{Sp}(#1; {\bf #2}) \else $\mathrm{Sp}(#1; {\bf #2})$\fi}
\def\usp#1{\ifmmode \mathrm{USp}(#1) \else $\mathrm{USp}(#1)$\fi}
\def\spin#1{\ifmmode \mathrm{Spin}(#1) \else $\mathrm{Spin}(#1)$\fi}
\def\su#1{\ifmmode \mathrm{SU}({#1}) \else $\mathrm{SU}(#1)$\fi}

\def\double #1{#1{\hbox{\kern-2pt $#1$}}}

\mathcode`\*="702A                  
\def\half{{\textstyle{1\over{\raise.1ex\hbox{$\scriptstyle{2}$}}}}}

\def \p{\partial}

\def \a{\alpha}

\def \b{\beta}

\def \d{\delta}

\def \g{\gamma}
\def\G{\Gamma}
\def\Gb{\overline\Gamma}
\def \l{\lambda}
\def\lh{\widehat\lambda}
\def\oh{\widehat\omega}

\def \L{\Lambda}

\def \o{\omega}

\def \t{\theta}

\def \r{\rho}

\def \s{\sigma}

\begin{document}

\begin{flushright}
\makebox[0pt][b]{}
\end{flushright}

\vspace{40pt}
\begin{center}
{\LARGE Relating the $b$ ghost and the vertex operators}

\vspace{14pt}

{\LARGE of the pure spinor superstring}

\vspace{40pt}
Osvaldo Chandia${}^{\spadesuit}$
\vspace{20pt}

{\em Departamento de Ciencias, Facultad de Artes Liberales
}\\
{\em Universidad Adolfo Ib\'a\~nez, Chile}\\

\vspace{40pt}
Brenno Carlini Vallilo${}^{\clubsuit}$
\vspace{20pt}

{\em Departamento de Ciencias F\'{\i}sicas, Facultad de Ciencias Exactas
}\\
{\em Universidad Andr\'es Bello, Chile}\\

\vspace{60pt}
{\bf Abstract}
\end{center}
The OPE between the composite $b$ ghost and the unintegrated vertex 
operator for massless states of the pure spinor superstring is 
computed and shown to reproduce the structure of the bosonic string result. 
The double pole vanishes in the Lorenz gauge and the 
single pole is shown to be equal to the corresponding integrated vertex operator.

\blfootnote{
${}^{\spadesuit}$ \href{mailto:ochandiaq@gmail.com}{ochandiaq@gmail.com}}
\blfootnote{
${}^{\clubsuit}$ \href{mailto:vallilo@gmail.com}{vallilo@gmail.com}}

\setcounter{page}0
\thispagestyle{empty}
\newpage

\tableofcontents

\parskip = 0.1in

\section{Introduction}

The pure spinor formalism for the superstring was introduced twenty
years ago by Berkovits \cite{Berkovits:2000fe}. It is an ad hoc method to quantize the
superstring in the Green-Schwarz formulation using a bosonic spinor
ghost satisfying the pure spinor constraint $\lambda
\gamma^m\lambda=0$. The formalism successfully describes the correct spectrum
\cite{Berkovits:2000nn,Berkovits:2001mx}
and scattering amplitudes\footnote{See {\em e.g.} \cite{Berkovits:2000ph,Berkovits:2005ng}.}
in flat space. The formalism was also used in curved spaces, in
particular $AdS_5\times S^5$ \cite{Mazzucato:2011jt} and its pp-wave
limit \cite{Berkovits:2002zv,Chandia:2018ekc}.

One of the most important challenges of the formalism is to explain
its origin from first principles. It is still not known the ungauged
fixed classical action that gives rise to the pure spinor ghosts and
the BRST charge. There are several interesting papers attacking this
problem  \cite{Matone:2002ft,Berkovits:2011gh,Berkovits:2014aia,Berkovits:2015yra,Jusinskas:2019vmd} but a
complete answer is still missing.  Related to this problem is the fact
that the usual reparametrization ghosts are composite operators in the
formalism. Consistency of the formalism requires the stress energy tensor to be 
BRST exact, however the existence of the $b$ ghost is a non-trivial fact.
It was first constructed in \cite{Berkovits:2001us,Berkovits:2004px}, 
and requires a picture changing operator multiplying the stress-energy tensor. Later, introducing non-minimal
variables, the $b$ ghost was understood as part of an underlying $N=2$
superconformal algebra \cite{Berkovits:2005bt,Berkovits:2006vi}. The
first example of a composite $b$ ghost in curved spaces was found in
\cite{Berkovits:2008ga} for the $AdS_5\times S^5$ background. The
construction was further simplified for flat spaces in
\cite{Berkovits:2013pla} which made possible the generalization for
on-shell heterotic backgrounds \cite{Chandia:2013ima,Berkovits:2014ama}. It is also
known how to define the $c$ ghost conjugate to $b$
\cite{Jusinskas:2013sha}.

In the case of the bosonic string the $b$ ghost plays a role in
relating integrated and unintegrated vertex operators
\begin{align}\label{fund1}
  b_{-1}U =V,
\end{align}
where $V$ is the integrated vertex operator and $U$ the unintegrated
one. We are also using the notation ${\cal O}_{n}{\cal A}$ meaning the
coefficient of the pole of order $n+h_{\cal O}$ in the OPE between the
two operators, with $h_{\cal O}$
being the conformal weight of $\cal O$. In the early days of the pure spinor
formalism the $b$ ghost was not known and the fundamental relation between the
types of vertices was in terms of the BRST charge
\begin{align}\label{fund2}
  QV=\partial U,
\end{align}
which is just a consequence of $Qb=T$. When $b$ is a fundamental
field (\ref{fund1}) is trivial to verify. However, in the pure
spinor formalism,  (\ref{fund2}) is much simpler from the computational
point of view. The relation (\ref{fund1}) was checked in
\cite{Grassi:2009fe} using the $b$ ghost \cite{Oda:2007ak} in
the so-called Y-formalism \cite{Oda:2005sd} that uses a constant pure
spinor to simplify many computations.
In this work we will show explicitly that the
relation (\ref{fund1}) works in the non minimal pure spinor formalism
without using the Y-formalism trick. The
calculation is rather intricate due to the many terms present in the
$b$ ghost. The computation also  requires a proper normal ordering
definition of the composite operators involved. One important difference
from \cite{Grassi:2009fe} is that (\ref{fund1}) is derived without
using Lorenz gauge, just as in the bosonic string case. For discussions
about the Siegel gauge in the context of pure spinor string
see  \cite{Aisaka:2009yp,Jusinskas:2015eza}.

This work is organized as follows. In section \ref{fengsiegelreview}
we review the gauge covariant description of massless state in open
bosonic string as discussed in \cite{Siegel:2003sv}. Section
\ref{review} contains a short review of the relevant aspects of the
pure spinor string and its $b$ ghost. We also list some useful
formulas used later. The computation of the OPE between the non
minimal $b$ and the unintegrated vertex operator is carried out in
section \ref{TheComputation}. The second order pole is computed first,
which vanishes in the Lorenz gauge. After it the computation of the
single pole is explained and it gives the expected integrated vertex
operator up to total derivatives and BRST exact terms. We end with
some comments and possible future work in section \ref{conclu}.

\section{The case of the bosonic string}
\label{fengsiegelreview}
In this section the work of \cite{Siegel:2003sv} is reviewed. After using the conformal gauge for the world-sheet metric, the bosonic string has a BRST charge given by
\begin{align}
Q=\oint c T_X + bc\p c,
\label{bos1}
\end{align}
where $T_X=-\frac12\p X_a \p X^a$ is the stress-energy of the $X$ field and the pair of odd variables $(b,c)$ are the Faddev-Popov ghost fields. The stress-energy tensor for these ghost fields is given by
\begin{align}
Qb=T_X+T_{bc}=-\frac12\p X_a \p X^a-2b\p c + c\p b ,
\label{bos101}
\end{align}
where $Qb$ is calculated according to
\begin{align}
Qb=\oint dw (c T_X + bc\p c)(w) b(z) ,
\label{bos102}
\end{align}
and the OPE's are obtained from the basic OPE's
\begin{align}
X^a(w) X^b(z) \to -\eta^{ab}\log |w-z|^2,\quad b(w) c(z) \to \frac1{(w-z)} .
\label{bos103}
\end{align}

The BRST charge is used to compute the space of the physical states of the bosonic string. They belong to the cohomology of $Q$, that is, any physical state is annihilated by $Q$ and two states differing in a $Q$-exact quantity represent the same physical state. The physical states of the bosonic string are represented by vertex operators of a given conformal dimension. Each state can be described in terms of unintegrated or integrated vertex operators depending on their role in the string scattering amplitudes. For the massless state, the unintegrated vertex operator is \cite{Siegel:2003sv}
\begin{align}
U=c\p X^a A_a - \frac12 \p c \p^a A_a .
\label{bos2}
\end{align}
Physical state condition gives the equation of motion of the field $A_a(X)$. In fact,
\begin{align}
QU=\frac12 c\p c \p X^a \p^b \p_{[b} A_{a]}=0\Rightarrow  \p^b \p_{[b} A_{a]}=0 ,
\label{bos3}
\end{align}
which is the equation of motion of a Maxwell field. The gauge invariance $\d U=Q\L$ gives the gauge transformation of the field $A_a(X)$. In fact,
\begin{align}
\d U=Q\L=c\p X^a \p_a \L-\frac12 \p c \Box \L =c\p X^a \d A_a - \frac12 \p c \p^a \d A_a \Rightarrow \d A_a=\p_a \L ,
\label{bos4}
\end{align}
where $\Box=\p^a\p_a$.  In these calculations we have used
\begin{align}
T_X(w) \p X^a A_a(z) \to &-\frac1{(w-z)^3} \p^a A_a(z) + \frac1{(w-z)^2} \left( \p X^a A_a(z)-\frac12\p X^a\Box A_a(z) \right) \cr &+ \frac1{(w-z)}\p\left( \p X^a A_a \right) ,
\label{bos5}
\end{align}
and
\begin{align}
T_X(w) f(X(z)) \to - \frac1{(w-z)^2} \frac12 \Box f(z) + \frac1{(w-z)} \p f(z) .
\label{bos6}
\end{align}
Using these results, the relation
\begin{align}
\p U = Q ( \p X^a A_a ) ,
\label{bos7}
\end{align}
is obtained. In this way, the integrated vertex operator, $V$, is defined such that $QV=\p U$. Therefore, the integrated vertex operator corresponding to the massless state is $V=\p X^a A_a$.

Let's compute the OPE between the stress-energy tensor $T$ and the unintegrated vertex operator $U$. It is given by
\begin{align}
T(w) U(z) \to -\frac1{(w-z)^2} \frac12 \left( c\p X^a \Box A_a - \frac12 \p c \Box \p^a A_a \right) +  \frac1{(w-z)} \p U .
\label{bos8}
\end{align}
Note that
\begin{align}
&Q (\p^a A_a) = \oint dw c(w) T_X(w) \p^a A_a = \oint dw c(w) \left( -\frac1{(w-z)^2} \frac12 \Box  (\p^a A_a) + \frac1{(w-z)} \p \p^a A_a \right) \cr
&=c \p X^a \p^b \p_a A_b-\frac12 \p c \Box  (\p^a A_a) =  c \p X^a  \Box A_a -\frac12 \p c \Box  (\p^a A_a) ,
\label{bos9}
\end{align}
where we first used (\ref{bos6}) and then (\ref{bos3}). Using (\ref{bos9}), the result of (\ref{bos8}) becomes
\begin{align}
T(w) U(z) \to \frac1{(w-z)^2} \left( -\frac12  Q (\p^a A_a) \right)+  \frac1{(w-z)} \p U .
\label{bos10}
\end{align}
This result has the expected single pole singularity and states that the double pole singularity is the BRST exact form of a function which is related to the Lorenz gauge for the gauge field $A_a$.
Let's compute now the OPE between $b(w)$ and $U$ of (\ref{bos2}). It turns out to be
\begin{align}
b(w) U(z) \to \frac1{(w-z)^2} \left( -\frac12 \p^a A_a \right) + \frac1{(w-z)} \left( \p X^a A_a \right),
\label{bos11}
\end{align}
where the single pole singularity is the integrated vertex operator and the vanishing of the double pole singularity is the Lorenz gauge for the gauge field.

Using the notation $b_n U$ being the pole of order $n+2$ of the OPE between $b$ and $U$, the OPE (\ref{bos11}) is equivalent to
\begin{align}
b_{-1}U=V ,
\label{bos12}
\end{align}
and the Lorenz gauge is
\begin{align}
b_0 U=0 .
\label{bos13}
\end{align}
Because $Qb=T$, acting with $Q$ the OPE (\ref{bos11}) gives the OPE (\ref{bos10}). In the next section the relation between unintegrated and integrated vertex operators through the existence of the $b$ ghost, as in (\ref{bos12}), will be generalized for the the pure spinor superstring.

\section{Review of the pure spinor superstring}
\label{review}

The basics of the pure spinor formulation of the superstring are
reviewed first, in particular the construction of the $b$ ghost
field is given.

The pure spinor string is given by conformal invariant system constructed out of the superspace variables in ten dimensions and pure spinor variables \cite{Berkovits:2000fe}.  The world sheet variables are $(X^a, \t^\a, p_\a, \l^\a, \o_\a)$, where $a=1, \dots, 10,\; \a=1, \dots, 16$, $p_\a$ is the conjugate variable of the odd superspace variable $\t^\a$, $\l^\a$ is the pure spinor variable which is an even variable constrained by $\l\g^a\l=0$, where $\g^a_{\a\b}$ are the $16\times 16$ symmetric $\g$ matrices in ten dimensions. The variable $\o_\a$ is the conjugate of $\l^\a$ and is defined up to $\d\o_\a=(\l\g^a)_\a \L_a$.

The quantization of the pure spinor superstring is given by the existence of the nilpotent charge
\begin{align}
Q=\oint \l^\a d_\a ,
\label{pure1}
\end{align}
where
\begin{align}
d_\a=p_\a-\frac12(\g^a\t)_\a\p X_a-\frac18(\g^a\t)_\a(\t\g_a\p\t) .
\label{pure2}
\end{align}
The charge $Q$ satisfies $Q^2=0$ because of the OPE
\begin{align}
d_\a(w) d_\b(z) \to -\frac1{(w-z)} \g^a_{\a\b} \Pi_a(z)
\label{pure3}
\end{align}
and the pure spinor condition. Here $\Pi_a=\p X_a +\frac12(\t\g_a\p\t)$.

Because nilpotency of $Q$, physical states of the pure spinor superstring are defined to be in the cohomology of $Q$. For the massless states, the corresponding unintegrated vertex operator is given by $U=\l^\a A_\a(X,\t)$. The superfield $A_\a$ satisfies the equations determined from $QU=0$. They are
\begin{align}
D_{(\a}A_{\b)}=\g_{\a\b}^a A_a,\quad D_\a A_a-\p_a A_\a=(\g_a)_{\a\b} W^\b,\quad D_\a W^\b=\frac14(\g^{ab})_\a{}^\b F_{ab} ,
\label{pure4}
\end{align}
where $D_\a=\p_\a+\frac12(\g^a)_\a\p_a$ is the covariant superspace derivative and $A_a, W^\a, F_{ab}=\p_{[a} A_{b]}$ are defined here. These definitions imply the equations of motion of super Maxwell in ten dimensions
\begin{align}
\p^b F_{ab}=\g^a_{\a\b} \p_a W^\b=0 ,
\label{pure5}
\end{align}
then the $\t^\a=0$ component of $A_a$ and $W^\a$ are the gauge field and the photino respectively. The gauge invariance comes from $\d U=Q\L$ which implies
\begin{align}
\d A_a=\p_a \L,\quad \d W^\a=0 ,
\label{pure6}
\end{align}
which give the gauge transformations of the photon and the photino.

The stress-energy tensor has vanishing central charge and it is given by
 \begin{align}
T=-\frac12 \p X_a \p X^a - p_\a \p\t^\a -\o_\a \p\l^\a .
\label{pure7}
\end{align}
The OPE between $T$ and $U$ is
\begin{align}
T(w) U(z) \to -\frac1{(w-z)^2} \frac12  \l^\a \Box A_\a + \frac1{(w-z)} \p U .
\label{pure8}
\end{align}
But \cite{Aisaka:2009yp}
\begin{align}
\Box A_\a = D_\a (\p^a A_a) .
\label{pure9}
\end{align}
This can be shown as follows
$$D_\a (\p^a A_a) = \p^a D_\a A_a = \p^a ( \p_a A_\a + (\g_a W)_\a ) = \Box A_a ,$$
because $W^\a$ satisfies the equation $\g^a_{\a\b} \p_a W^\b=0$. 
The equation \ref{pure9}  imply that
\begin{align}
\l^\a \Box A_\a = Q (\p^a A_a) .
\label{pure12}
\end{align}
Therefore, the OPE (\ref{pure8}) can be written as
\begin{align}
T(w) U(z) \to -\frac1{(w-z)^2} \frac12 Q (\p^a A_a)  + \frac1{(w-z)} \p U ,
\label{pure13}
\end{align}
which has the form (\ref{bos10}) of the bosonic string.

To obtain an expression similar to (\ref{bos11}) a $b$ ghost field is necessary. It is not known how to gauge-fix a local symmetry and produce these type of ghosts. But there exists an odd variable of conformal dimension two which satisfied $Qb=T$. It is necessary to add the so called non-minimal pure spinor variables to reach this goal \cite{Berkovits:2005bt}. They are the pair of even conjugate variables $(\lh_\a, \oh^\a)$ and the pair of odd conjugate variables $(r_\a, s^\a)$ which are constrained to satisfy
\begin{align}
(\lh\g^a\lh)=(r\g^a\lh)=0 ,
\label{pure14}
\end{align}
and the variables $\oh^\a$ and $s^\a$ are defined up to
\begin{align}
\d s^\a=(\g^a\lh)^\a {\widetilde\L}_a ,\quad \d \oh^\a=(\g^a\lh)^\a {\widetilde{\widetilde\L}}_a-(\g^a r)^\a  {\widetilde\L}_a .
\label{pure15}
\end{align}
The $b$ ghost turns out to be \cite{Berkovits:2013pla}
\begin{align}
b=-s^\a\p\lh_\a+\Pi^a \Gb_a - \frac1{4(\l\lh)} (\l\g^{ab}r)\Gb_a\Gb_b-\frac1{4(\l\lh)} \left( J (\lh\p\t)+N^{ab}(\lh\g_{ab}\p\t) \right),
\label{pure16}
\end{align}
where $J=-\l^\a\o_\a$, $N^{ab}=\frac12(\l\g^{ab}\o)$ and
\begin{align}
\Gb_a=\frac1{2(\l\lh)} (d\g_a\lh)+\frac1{8(\l\lh)^2}(r\g_{abc}\lh) N^{bc} .
\label{pure17}
\end{align}
All the products here are in normal order which is defined below. The calculation of $Qb=T$ can be found in \cite{Berkovits:2014ama}. We are interested in the OPE $b(w) U(z)$ to obtain a result similar to (\ref{bos11}) of the bosonic case. For this purpose it is useful to expand the $b$ ghost in powers of the odd non-minimal pure spinor variable $r_\a$ as
\begin{align}
b=-s^\a\p\lh_\a+b^{(0)}+b^{(1)}+b^{(2)}+b^{(3)} ,
\label{pure18}
\end{align}
where $b^{(n)}$ is the term with $n$ factors of the non-minimal variable $r_\a$ in the $b$ ghost. They are given by
\begin{align}
&b^{(0)}=-\left( \frac1{4(\l\lh)}\lh_\a ~ J \p\t^\a \right)  - \left(  \frac1{4(\l\lh)}(\lh\g_{ab})_\a ~ N^{ab} \p\t^\a \right) + \left(\frac1{2(\l\lh)} (\g^a\lh)^\a ~\left( \Pi_a ~ d_\a \right) \right)  ,\cr
&b^{(1)}= \left( \frac1{8(\l\lh)^2} (r\g^{abc}\lh) ~ \left( N_{bc} ~ \Pi_a \right) \right) - \left( \frac1{16(\l\lh)^3} (\l\g^{ab}r) (\g_a\lh)^\a (\g_b\lh)^\b ~ \left( d_\a ~ d_\b \right) \right)  ,\cr
&b^{(2)}=\left(\frac1{64(\l\lh)^4} (\l\g^{ab}r) (\g_{[a}\lh)^\a (r\g_{b]cd}\lh) ~ \left( N^{cd} ~ d_\a \right) \right),\cr
&b^{(3)}= - \left( \frac1{256(\l\lh)^5} (\l\g^{ab}r) (r\g_{acd}\lh) (r\g_{bef}\lh) ~ \left( N^{cd} ~ N^{ef} \right) \right) .
\label{pure19}
\end{align}
Note that the term with $(\l\lh)^{-3}$ in $b^{(1)}$ is usually written as being proportional to $(\l\lh)^{-2} (\lh \g^{abc} r) \left( (d\g_{abc}d) \right)$. To prove that both terms are equal one first uses that $\left( d_\a d_\b \right)=\frac1{96} \g^{abc}_{\a\b} \left( (d\g_{abc}d) \right)$  in $b^{(1)}$. Then, after using the identities $(\g_a\lh)^\a(\g^a r)^\b=-(\g_a r)^\a (\g^a\lh)^b$  and $(\g_a\lh)^\a(\g^a\lh)^\b=0$, which are consequences of the pure spinor constraints (\ref{pure14}), one obtains the usual expression in $b^{(1)}$. A similar procedure is used in $b^{(3)}$ to have a term with $(\l\lh)^{-4}$ instead of the term shown in (\ref{pure19}). 

The normal oder of two operators in (\ref{pure19}) is defined as
\begin{align}
\left( A ~ B \right) (z)=\oint \frac1{(y-z)} A(y) B(z).
\label{pure190}
\end{align}
To compute the OPE $b(w) U(z)$ we obtain first some useful OPE's in the next subsection. See \cite{Jusinskas:2013yca} and \cite{Sepulveda:2020wwq}. In particular, we will use the identities
\begin{align}
&\left( A ~ B \right) = (-1)^{ab} \left( B ~ A \right) + \left( [ A, B ] \right) ,\cr
&\left( A ~ \left( B ~ C \right) \right) = (-1)^{ab}  \left( B ~
  \left( A ~ C \right) \right) + \left( \left( [ A , B ] \right) ~ C \right) ,
\label{p34}
\end{align}
where $a$ and $b$ are the Grassmann parities of $A$ and $B$ respectively and $[A,B]=AB-(-1)^{ab} BA$. It turns out that $\left( [A,B] \right) = \p \Delta$ in the case where  the OPE $A(w) B(z)$ has the form $(w-z)^{-1} \Delta(z)$.

\subsection{Useful OPE's}
\label{ope}

Consider the OPE $\left( \Pi_a ~ d_\a \right)(w) A_\b(z)$ which is necessary in $b^{(0)}(w) U(z)$. Recall
\begin{align}
\left( \Pi_a ~ d_\a \right) (w) = \oint_{\G_w} dy \frac1{(y-w)} \Pi_a(y) d_\a(w) ,
\label{pure20}
\end{align}
where $\G_w$ encircles counterclockwise the point $w$. Then,
\begin{align}
\left( \Pi_a ~ d_\a \right) (w)  A_\b(z)=\oint_{\G_w} dy \frac1{(y-w)} \Pi_a(y) d_\a(w) A_\b(z) .
\label{pure21}
\end{align}
By deforming the path $\G_w$ such that it is equal to $\G_{wz}-\G_z$, where $\G_{wz}$ encircles counterclockwise both points $w$ and $z$, and $\G_z$ encircles counterclockwise $z$ but not $w$ (see the figure below), one obtains
\begin{align}
\left( \Pi_a ~ d_\a \right) (w)  A_\b(z) = \oint_{\G_{wz}} dy \frac1{(y-w)} \Pi_a(y) d_\a(w) A_\b(z) - \oint_{\G_z}  dy  \frac1{(y-w)} d_\a(w) \Pi_a(y) A_\b(z) .
\label{pure22}
\end{align}

\hspace{2cm}\includegraphics[scale=0.5]{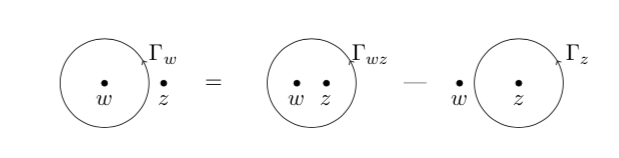}

\noindent
For the first integral one expands in $(w-z)$ and for the second integral one expands in $(y-z)$. Both expansions include singularities leading to
\begin{align}
\left( \Pi_a ~ d_\a \right) (w) A_\b(z) &= \oint_{\G_{wz}} dy \frac1{(y-w)} \Pi_a(y) \left( \frac1{(w-z)} D_\a A_\b(z) + {\cal O}(w-z)^0 \right) \cr
&-d_\a(w) \oint_{\G_z}  dy  \frac1{(y-w)} \left( -\frac1{(y-z)} \p_a A_\b(z) + {\cal O}(y-z)^0   \right) .
\label{pure23}
\end{align}
Note that $\Pi_a(y)$ produces singularities in $(y-z)$ when approaches the operators defined in the point $z$ in the first line. But these contribution will vanish because
\begin{align}
\oint _{\G_{wz}} dy \frac1{(y-w)(y-z)}=\frac1{(w-z)} \oint _{\G_{wz}} dy \left(  \frac1{(y-w)} - \frac1{(y-z)}  \right) = 0  .
\label{pure24}
\end{align}
Higher order poles will also vanish because
\begin{align}
\oint _{\G_{wz}} dy \frac1{(y-w)(y-z)^n}= 0 ,
\label{pure25}
\end{align}
as can be obtained from (\ref{pure24}) by taking derivatives with respect to $z$.  For the second integral in (\ref{pure23}),
\begin{align}
\oint _{\G_{z}} dy \frac1{(y-w)(y-z)}=\frac1{(w-z)} \oint _{\G_{z}} dy \left(  \frac1{(y-w)} - \frac1{(y-z)}  \right) = -\frac1{(w-z)}  .
\label{pure26}
\end{align}
Then, (\ref{pure23}), as $w \to z$, is
\begin{align}
\left( \Pi_a ~ d_\a \right) (w) A_\b(z)  \to \frac1{(w-z)} \Pi_a D_\a A_\b - \frac1{(w-z)} d_\a(w) \p_a A_\b(z) .
\label{pure27}
\end{align}
For the second term here the non-singular term in the OPE $d_\a(w)$ with a superfield $\Psi(X,\t)(z)$ as $w\to z$. This is given by
\begin{align}
d_\a(w) \Psi(z) \to \frac1{(w-z)} D_\a\Psi(z) + \left( d_\a ~ \Psi \right)(z) .
\label{pure28}
\end{align}
Finally, using (\ref{pure28}) one obtains
\begin{align}
\left( \Pi_a ~ d_\a \right) (w) A_\b(z) &\to \frac1{(w-z)^2} \left( -D_\a \p_a A_\b(z)  \right) \cr
&+ \frac1{(w-z)} \left( \left( \Pi_a ~ D_\a A_\b \right)(z) - \left( d_\a ~ \p_a  A_\b \right)(z)   \right)
\label{pure29}
\end{align}
Following a similar procedure one obtains
\begin{align}
\left( N_{bc} ~ \Pi_a \right) (w) \l^\a A_\a (z) &\to\frac1{(w-z)^2} \frac12 (\l\g_{bc})^\a \p_a A_\a(z)\cr
&+\frac1{(w-z)}\left( -\left( N_{bc} ~ \l^\a\p_a A_\a \right) - \frac12 \left( \Pi_a ~ (\l\g_{bc})^\a A_\a \right)   \right) ,
\label{pure299}
\end{align}
\begin{align}
\left( d_\a d_\b \right) (w) A_\g (z) &\to \frac1{(w-z)^2} \left( \frac12 D_{[\a} D_{\b]} A_\g(z) \right) \cr
&+ \frac1{(w-z)} \left( d_{[\a} D_{\b]} ~ A_\g(z)  \right) ,
\label{pure30}
\end{align}
\begin{align}
\left( N^{cd} ~ d_\a \right) (w) \l^\b A_\b(z)  &\to  \frac1{(w-z)^2}  \left(-\frac12 (\l\g^{cd})^\b D_\a A_\b \right) \cr
& + \frac1{(w-z)} \left(  \left( N^{cd} ~ \l^\b D_\a A_\b \right) - \frac12 \left( d_\a ~ (\l\g^{cd})^\b A_\b \right) \right).
\label{pure31}
\end{align}
\begin{align}
\left( N^{cd} ~ N^{ef} \right) (w) \l^\a(z) \to &\frac1{(w-z)^2} \frac18 \left(\l\left( \g^{cd}\g^{ef}+\g^{ef}\g^{cd}\right)\right)^\a \cr
+&\frac1{(w-z)} \left( -\frac12 \left( N^{cd} ~ (\l\g^{ef})^\a \right) - \frac12 \left( N^{ef} ~ (\l\g^{cd})^\a \right) \right) .
\label{pure32}
\end{align}

\section{Computation of $b(w) U(z)$}
\label{TheComputation}

We now compute the OPE  $b(w) U(z)$ using the above results. First we will consider $b^{(0)}$. The first two terms in (\ref{pure19})  contribute to the OPE with $U(z)$ with
\begin{align}
 \frac1{(w-z)} \left( -\frac1{4(\l\lh)} \left( (\lh\p\t)(\l A) -\frac12 (\lh\g_{ab}\p\t) (\l\g^{ab})^\a A_\a \right) \right)  .
\label{pure320}
\end{align}
After using (\ref{pure29}), the last term of  $b^{(0)}$ leads to
\begin{align}
&\oint_{\G_w} \frac{dy}{(y-w)} ~ \frac1{2(\l\lh)} (\g^a\lh)^\a(y) \l^\b(z) \left( \Pi_a ~ d_\a \right) (w) A_\b(z) \cr
&=\oint_{\G_w} \frac{dy}{(y-w)} ~ \frac1{2(\l\lh)} (\g^a\lh)^\a(y) \l^\b(z) ~ \frac1{(w-z)^2} \left( -D_\a \p_a A_\b (z) \right) \cr
&+\oint_{\G_w} \frac{dy}{(y-w)} ~ \frac1{2(\l\lh)} (\g^a\lh)^\a(y) ~ \frac1{(w-z)}\left(  \left( \Pi_a ~ \l^\b D_\a A_\b \right)(z) - \left( d_\a ~ \l^\b \p_a  A_\b \right)(z)   \right) \cr
&=\frac1{(w-z)^2} \left(   -\frac1{2(\l\lh)} (\g^a\lh)^\a \l^\b \p_a D_\a A_\b \right) (z) \cr
&+\frac1{(w-z)} \left( \frac1{2(\l\lh)} (\g^a\lh)^\a ~ \left(  \left( \Pi_a ~ \l^\b D_\a A_\b \right) \right) \right) (z) \cr
&- \frac1{(w-z)} \left( \frac1{2(\l\lh)} (\g^a\lh)^\a ~ \left(  \left( d_\a ~ \l^\b \p_a A_\b \right) \right) \right) (z) \cr
&-\frac1{(w-z)} \p\left( \frac1{2(\l\lh)} (\g^a\lh)^\a \right)  \l^\b D_\a \p_a A_\b (z)
\label{pure321}
\end{align}
Using the results of (\ref{pure320}) and (\ref{pure321}) we obtain
\begin{align}
b_0^{(0)} U =  -\frac1{2(\l\lh)} (\g^a\lh)^\a \l^\b \p_a D_\a A_\b ,
\label{pure322}
\end{align}
and
\begin{align}
b_{-1}^{(0)} U &=  -\frac1{4(\l\lh)} \left( (\lh\p\t)(\l A) -\frac12 (\lh\g_{ab}\p\t) (\l\g^{ab})^\a A_\a \right) - \p\left( \frac1{2(\l\lh)} (\g^a\lh)^\a \right) \l^\b D_\a \p_a A_\b \cr
&+ \left( \frac1{2(\l\lh)} (\g^a\lh)^\a ~ \left( \Pi_a ~ \l^\b D_\a A_\b \right) \right) - \left( \frac1{2(\l\lh)} (\g^a\lh)^\a ~  \left( d_\a ~ \l^\b \p_a A_\b  \right) \right).
\label{pure323}
\end{align}

Now we calculate the contributions coming from $b^{(1)}$. After using (\ref{pure299}) and (\ref{pure30}) , the OPE with $U$ is
\begin{align}
b^{(1)}(w) U(z) &\to \oint_{\G_w} \frac{dy}{(y-w)} ~ \frac1{8(\l\lh)^2} (r\g^{abc}\lh) (y) ~ \frac1{(w-z)^2} \frac12 (\l\g_{bc})^\a \p_a A_\a \cr
&-\oint_{\G_w} \frac{dy}{(y-w)} ~ \frac1{8(\l\lh)^2} (r\g^{abc}\lh) (y) ~  \frac1{(w-z)} \left( N_{bc} ~ \l^\a \p_a A_\a \right)  \cr
&-\oint_{\G_w} \frac{dy}{(y-w)} ~ \frac1{8(\l\lh)^2} (r\g^{abc}\lh) (y) ~ \frac1{(w-z)} \frac12 \left( \Pi_a ~ (\l\g_{bc})^\a A_\a \right) \cr
-\oint_{\G_w} & \frac{dy}{(y-w)} ~ \frac1{16(\l\lh)^3} (\l\g^{ab}r)  (\g_a\lh)^\a (\g_b\lh)^\b (y) \l^\g(z)  ~ \frac1{(w-z)^2} \frac12 D_{[\a} D_{\b]} A_\g (z) \cr
-\oint_{\G_w} & \frac{dy}{(y-w)} ~ \frac1{16(\l\lh)^3} (\l\g^{ab}r)  (\g_a\lh)^\a (\g_b\lh)^\b (y) \l^\g(z)  ~ \frac1{(w-z)} \left( d_{[\a} ~ D_{\b]} A_\g (z) \right) ,
\label{pure460}
\end{align}
which implies
\begin{align}
b_0^{(1)} U = \frac1{16(\l\lh)^2} (r\g^{abc}\lh) (\l\g_{bc})^\a \p_a A_\a - \frac1{16(\l\lh)^3} (\l\g^{ab}r)(\g_a\lh)^\a(\g_b\lh)^\b \l^\g D_\a D_\b A_\g ,
\label{pure461}
\end{align}
and
\begin{align}
b_{-1}^{(1)} U &= -\left( \frac1{8(\l\lh)^2} (r\g^{abc}\lh) ~ \left( N_{bc} ~ \l^\a \p_a A_\a \right) \right) \cr
&- \left( \frac1{16(\l\lh)^2} (r\g^{abc}\lh) ~ \left( \Pi_a ~ (\l\g_{bc})^\a A_\a \right) \right) \cr
&-\left(  \frac1{8(\l\lh)^3} (\l\g^{ab}r)  (\g_a\lh)^\a (\g_b\lh)^\b ~ \left( d_\a ~ \l^\g D_\b A_\g \right) \right) \cr
&+ \p\left( \frac1{16(\l\lh)^2} (r\g^{abc}\lh) \right) (\l\g_{bc})^\a \p_a A_\a  \cr
&-\p\left( \frac1{16(\l\lh)^3} (\l\g^{ab}r)  (\g_a\lh)^\a (\g_b\lh)^\b \right)  \l^\g D_\a D_\b A_\g .
\label{pure462}
\end{align}

Next we focus on the terms coming from $b^{(2)}$.  After  using (\ref{pure31}), the OPE with $U$ is
\begin{align}
b^{(2)}(w) U(z) &\to \oint_{\G_w} \frac{dy}{(y-w)}   \frac1{32(\l\lh)^4}  (\l\g^{ab}r) (r\g_{bcd}\lh) (\g_a\lh)^\a (y) ~ \frac1{(w-z)^2} \left( -\frac12 (\l\g^{cd})^\b D_\a A_\b \right) (z) \cr
&+\oint_{\G_w} \frac{dy}{(y-w)}   \frac1{32(\l\lh)^4}  (\l\g^{ab}r) (r\g_{bcd}\lh) (\g_a\lh)^\a (y) ~  \frac1{(w-z)} \left( N^{cd} ~ \l^\b D_\a A_\b \right) (z) \cr
&-\oint_{\G_w} \frac{dy}{(y-w)}   \frac1{32(\l\lh)^4}  (\l\g^{ab}r) (r\g_{bcd}\lh) (\g_a\lh)^\a (y) ~  \frac1{(w-z)} \frac12 \left( d_\a ~ (\l\g^{cd})^\b A_\b \right) (z) ,
\label{pure620}
\end{align}
which implies
\begin{align}
b_0^{(2)} U = -\frac1{64(\l\lh)^4}  (\l\g^{ab}r) (r\g_{bcd}\lh) (\g_a\lh)^\a  (\l\g^{cd})^\b D_\a A_\b ,
\label{pure621}
\end{align}
and
\begin{align}
b_{-1}^{(2)} U &= \left( \frac1{32(\l\lh)^4}  (\l\g^{ab}r) (r\g_{bcd}\lh) (\g_a\lh)^\a ~ \left( N^{cd} ~ \l^\b D_\a A_\b \right) \right)  \cr
&- \left( \frac1{64(\l\lh)^4}  (\l\g^{ab}r) (r\g_{bcd}\lh) (\g_a\lh)^\a ~ \left( d_\a ~ (\l\g^{cd})^\b A_\b \right) \right) \cr
&-\p\left( \frac1{64(\l\lh)^4}  (\l\g^{ab}r) (r\g_{bcd}\lh) (\g_a\lh)^\a \right)   (\l\g^{cd})^\b D_\a A_\b .
\label{pure622}
\end{align}

Finally, after using (\ref{pure32}), the OPE between $b^{(3)}$ and $U$ is
\begin{align}
b^{(3)}(w) U(z) \to& -\oint_{\G_w} \frac{dy}{(y-w)} \frac1{1024(\l\lh)^5} (\l\g^{ab}r) (r\g_{acd}\lh) (r\g_{bef}\lh) (y) ~ \frac1{(w-z)^2} (\l\g^{cd}\g^{ef})^\a A_\a(z) \cr
&+\oint_{\G_w} \frac{dy}{(y-w)} \frac1{256(\l\lh)^5} (\l\g^{ab}r) (r\g_{acd}\lh) (r\g_{bef}\lh) (y) ~  \frac1{(w-z)} \left( N^{cd} ~ (\l\g^{ef})^\a A_\a \right) (z) ,
\label{pure8110}
\end{align}
which implies
\begin{align}
b_0^{(3)} U = -\frac1{1024(\l\lh)^5} (\l\g^{ab}r) (r\g_{acd}\lh) (r\g_{bef}\lh)  (\l\g^{cd}\g^{ef})^\a A_\a ,
\label{pure8111}
\end{align}
and
\begin{align}
b_{-1}^{(3)} U &= \left(  \frac1{256(\l\lh)^5} (\l\g^{ab}r) (r\g_{acd}\lh) (r\g_{bef}\lh) ~ \left( N^{cd} ~ \left( (\l\g^{ef})^\a A_\a \right) \right) \right) \cr
&-\p\left( \frac1{1024(\l\lh)^5} (\l\g^{ab}r) (r\g_{acd}\lh) (r\g_{bef}\lh)  \right) ~  (\l\g^{cd}\g^{ef})^\a A_\a .
\label{pure8112}
\end{align}

\subsection{$b_0 U$ }

We now collect and simplify the results of the previous subsection for $b_0 U$. Let's simplify the results of (\ref{pure322}), (\ref{pure461}), (\ref{pure621}) and (\ref{pure8111}). The first of these results is
\begin{align}
b_0^{(0)} U =  -\frac1{2(\l\lh)} (\g^a\lh)^\a \l^\b \p_a D_\a A_\b .
\label{p1}
\end{align}
Using (\ref{pure4}) this expression becomes
\begin{align}
b_0^{(0)} U = -\frac12 \p \cdot A - \frac1{4(\l\lh)} (\lh\g^{ab}\l) F_{ab} + \frac1{2(\l\lh)} (\g^a\lh)^\a Q ( \p_a A_\a ) ,
\label{p2}
\end{align}
for the second term we can use the fact that $(\g^{ab}\l)^\a F_{ab}$ is equal to $-4 Q W^\a$ so that
\begin{align}
b_0^{(0)} U = -\frac12 \p \cdot A + \frac1{(\l\lh)} \lh_\a Q W^\a  + \frac1{2(\l\lh)} (\g^a\lh)^\a Q ( \p_a A_\a ) .
\label{p3}
\end{align}
Recall  (\ref{pure461}),
\begin{align}
b_0^{(1)} U = \frac1{16(\l\lh)^2} (r\g^{abc}\lh) (\l\g_{bc})^\a \p_a A_\a - \frac1{16(\l\lh)^3} (\l\g^{ab}r)(\g_a\lh)^\a(\g_b\lh)^\b \l^\g D_\a D_\b A_\g .
\label{p03}
\end{align}
Using the identity
\begin{align}
\d_\g^\a \d_\b^\r + \frac12 (\g^{ab})_\g{}^\a (\g^{ab})_\b{}^\r - 2 \g_a^{\a\r} \g^a_{\b\g} = -4\d_\b^\a \d_\g^\r ,
\label{pure41}
\end{align}
and the pure spinor conditions for the non-minimal variables in the first term and anticommuting $D_\a$ with $D_\b$ together with the equations  (\ref{pure4}) one obtains
\begin{align}
b_0^{(1)} U &= \left( -\frac1{2(\l\lh)} (\g^a r)^\a + \frac1{4(\l\lh)^2} (\l\g^b\g^a r) (\g_b\lh)^\a \right) \p_a A_\a \cr
&-\frac1{8(\l\lh)^3} (\l\g^{ab}r) (\lh\g_b\g^c\l) (\g_a\lh)^\a \p_c A_\a \cr
& - \frac1{16(\l\lh)^3} (\l\g^{ab}r)(\lh\g_b\g^c\l) (\lh\g_a\g_c W)    \cr
&- \frac1{16(\l\lh)^3} (\l\g^{ab}r)(\g_a\lh)^\a(\g_b\lh)^\b Q ( D_\a A_\b ) \cr
&=\left( -\frac1{2(\l\lh)} (\g^a r)^\a + \frac1{2(\l\lh)^2} (\l r) (\g^a\lh)^\a \right) \p_a A_\a  \cr
&+\left( -\frac1{(\l\lh)} r_\a + \frac1{(\l\lh)^2} (\l r) \lh_\a \right) W^\a \cr
&- \frac1{16(\l\lh)^3} (\l\g^{ab}r)(\g_a\lh)^\a(\g_b\lh)^\b Q ( D_\a A_\b ) ,
\label{p4}
\end{align}
then,
\begin{align}
b_0^{(1)} U&= Q\left( \frac1{2(\l\lh)} (\g^a\lh)^\a \right) \p_a A_\a + Q \left( \frac1{(\l\lh)} \lh_\a \right) W^\a \cr
&- \frac1{16(\l\lh)^3} (\l\g^{ab}r)(\g_a\lh)^\a(\g_b\lh)^\b Q ( D_\a A_\b ) .
\label{p5}
\end{align}
Recall now (\ref{pure621}),
\begin{align}
b_0^{(2)} U = -\frac1{64(\l\lh)^4}  (\l\g^{ab}r) (r\g_{bcd}\lh) (\g_a\lh)^\a  (\l\g^{cd})^\b D_\a A_\b ,
\label{p6}
\end{align}
Using (\ref{pure41}) and the pure spinor conditions this expression becomes
\begin{align}
b_0^{(2)} U &= \frac1{8(\l\lh)^3}  (\l\g^{ab}r) (\g_a\lh)^\a (\g_b r)^\b D_\a A_\b \cr
&+\frac3{16(\l\lh)^4} (\l r) (\l \g^{ab} r) (\g_a \lh)^\a (\g_b\lh)^\b D_\a A_\b ,
\label{p6}
\end{align}
in the first term we use $D_\a A_\b = \frac12 D_{(\a} A_{\b)} + \frac12 D_{[\a} A_{\b]}$ and the symmetric part does not contribute because of the pure spinor conditions, so
\begin{align}
b_0^{(2)} U &= \frac1{16(\l\lh)^3}  (\l\g^{ab}r) \left( (\g_a\lh)^\a (\g_b r)^\b + (\g_a r)^\a (\g_b \lh)^\b \right) D_\a A_\b \cr
&+\frac3{16(\l\lh)^4} (\l r) (\l \g^{ab} r) (\g_a \lh)^\a (\g_b\lh)^\b D_\a A_\b \cr
&=Q\left( \frac1{16(\l\lh)^3} (\l\g^{ab}r) (\g_a\lh)^\a (\g_b \lh)^\b \right) D_\a A_\b .
\label{p7}
\end{align}
Finally, recall (\ref{pure8111}),
\begin{align}
b_0^{(3)} U = -\frac1{1024(\l\lh)^5} (\l\g^{ab}r) (r\g_{acd}\lh) (r\g_{bef}\lh)  (\l\g^{cd}\g^{ef})^\a A_\a ,
\label{p8}
\end{align}
which vanishes. In fact, this contains the factor
\begin{align}
(\l\g^{ab}r) (r\g_{acd}\lh) (r\g_{bef}\lh)(\l\g^{cd}\g^{ef})^\a=(\l\g^{ab}r) (r\g_{acd}\lh) (r\g_b)^\b \lh_\g (\g_{ef})_\b{}^\g (\g^{ef})_\r{}^\a (\l\g^{cd})^\r ,
\label{p9}
\end{align}
using the identity (\ref{pure41}) this is equal to
\begin{align}
-8(\l\g^{ab}r) (r\g_{acd}\lh) (\l\g^{cd}\lh) (\g_b r)^\a + 4 (\l\g^{ab}r) (r\g_{acd}\lh)(\l\g^{cd}\g^e\g_b r) (\g_e \lh )^\a .
\label{p10}
\end{align}
The first term vanishes because $(\lh\g_a)^\a(\lh\g^a)^\b=0$. The second term is proportional to
\begin{align}
&(\l\g^{ab}r)(r\g_a)^\b\lh_\g \l^\r (\g^e \g_b r)_\s (\g_{cd})_\b{}^\g (\g^{cd})_\r{}^\s (\g_e\lh)^\a = 8 (\l\lh) (\l\g^{ab}r)(r\g_{abe}r) (\g^e\lh)^\a \cr
&=8 (\l\lh) \l^\b r_\g r_\r (\g_e r)^\s (\g^{ab})_\b{}^\g (\g_{ab})_\s{}^\r (\g^e\lh)^\a = 0 ,
\label{p11}
\end{align}
because (\ref{pure41}). Therefore,
\begin{align}
b_0^{(3)} U = 0 .
\label{p12}
\end{align}
Therefore, adding the results of (\ref{p3}), (\ref{p5}), (\ref{p7}) and (\ref{p12}) we obtain, up to a BRST exact term that
\begin{align}
b_0 U = -\frac12 \p \cdot A ,
\label{p13}
\end{align}
which is equal to the bosonic string case of (\ref{bos11}).

\subsection{$b_{-1} U$}

We now collect and simplify the results of the previous subsection for $b_{-1} U$. We will consider first the the results of (\ref{pure323}), (\ref{pure462}), (\ref{pure622}) and (\ref{pure8112}). The first of these results is
\begin{align}
b_{-1}^{(0)} U &=  -\frac1{4(\l\lh)} \left( (\lh\p\t)(\l A) -\frac12 (\lh\g_{ab}\p\t) (\l\g^{ab})^\a A_\a \right) - \p\left(\frac1{2(\l\lh)} (\g^a\lh)^\a \right) \l^\b D_\a \p_a A_\b \cr
&+\left( \frac1{2(\l\lh)} (\g^a\lh)^\a ~   \left( \Pi_a ~ \l^\b D_\a A_\b \right) \right) - \left( \frac1{2(\l\lh)} (\g^a\lh)^\a ~ \left(  d_\a ~ \l^\b \p_a A_\b \right) \right) .
\label{p14}
\end{align}
In the first term of the second line we use (\ref{pure4}) so that this term is equal to
\begin{align}
&\left( \frac1{2(\l\lh)} (\g^a\lh)^\a ~ \left( \Pi_a ~ (\l\g^b)_\a A_b \right) \right) -  \left( \frac1{2(\l\lh)} (\g^a\lh)^\a ~ \left( \Pi_a ~ Q A_\a \right) \right) \cr
=& \frac1{2(\l\lh)} (\lh\g^a\g^b\l) \Pi_a A_b + \frac1{2(\l\lh)} (\g^a\lh)^\a (\l\g_a\p\t) A_\a  - \left( \frac1{2(\l\lh)} (\g^a\lh)^\a ~  Q\left( \Pi_a ~ A_\a \right) \right) ,
\label{p15}
\end{align}
the second term here adds to the first term in (\ref{p14}) to give $\p\t^\a A_\a$ which is one of the terms of the integrated vertex operator. Consider the last term in
(\ref{p14}), again using (\ref{pure4}) it is equal to
\begin{align}
&\frac1{2(\l\lh)} (\lh\g^b\g^a\l) \Pi_a A_b + \left( \frac1{2(\l\lh)} (\g^a\lh)^\a ~  Q\left( d_\a ~ A_a \right) \right) + \left( \frac1{2(\l\lh)} (\g^a\lh)^\a ~  \left( d_\a ~ (\l\g_a W) \right) \right) \cr
&-\frac1{2(\l\lh)} (\lh\p\l) \p \cdot A + \frac1{4(\l\lh)} (\lh\g^{ab}\p\l) F_{ab} ,
\label{p16}
\end{align}
where total derivative terms were ignored. The first term here adds to the first term in (\ref{p15}) to give $\Pi^a A_a$ which is also part of the integrated vertex operator. In the last term of the first line in (\ref{p16}) we can use the identity (\ref{pure41}) so that this term is equal to
\begin{align}
d_\a W^\a +\left(  (\l d) ~~ \frac1{4(\l\lh)} \lh_\a W^\a \right) - \left( \frac12(\l\g^{ab} d) ~~ \frac1{4(\l\lh)} (\lh\g_{ab}W) \right) ,
\label{p17}
\end{align}
and we have obtained a third term of the integrated vertex operator. Using $QJ=-(\l d)$ and $QN^{ab}=\frac12(\l\g^{ab}d)$ together with (\ref{p34}) this expression becomes
\begin{align}
&d_\a W^\a + \frac12 N^{ab} F_{ab} + \left( \frac1{4(\l\lh)^2} (\lh\p\l) \lh_\a \right) Q W^\a \cr
& -   \left(  ( J W^\a + N^{ab} (\g_{ab} W)^\a ) ~~ Q \left( \frac1{4(\l\lh)} \lh_\a \right) \right) ,
\label{p170}
\end{align}
where the identity \cite{Berkovits:2002qx}
\begin{align}
\left( N_{ab} ~(\l\g^b)_\a \right) = -\frac12 \left( J ~ (\l\g_a)_\a\right) - 2 (\p\l\g_a)_\a ,
\label{p20}
\end{align}
was used. It remains to consider the second term in the first line of (\ref{p14}), it is equal to
\begin{align}
\p\left( \frac1{2(\l\lh)} (\g^a\lh)^\a\right) Q\left( \p_a A_\a
  \right)  - \frac1{(\l\lh)} \lh_\a \p \left( Q W^\a \right)  + \frac1{2(\l\lh)} (\lh\p\l) \p \cdot A .
\label{p24}
\end{align}
Adding all the contributions we obtain
\begin{align}
b_{-1}^{(0)} U &= \p\t^\a A_\a + \Pi^a A_a + d_\a W^\a + \frac12 N^{ab} F_{ab} \cr
&+\left( \frac1{2(\l\lh)} (\g^a\lh)^\a ~  Q \left( \left( d_\a ~ A_a \right) - \left( \Pi_a ~ A_\a \right)  \right) \right) \cr
& -\left(  ( J W^\a + N^{ab} (\g_{ab} W)^\a ) ~~ Q \left( \frac1{4(\l\lh)} \lh_\a \right) \right)+\p\left( \frac1{2(\l\lh)} (\g^a\lh)^\a\right) Q\left( \p_a A_\a \right) \cr
&- \frac1{(\l\lh)} \lh_\a \p \left( Q W^\a \right)  + \left( \frac1{(\l\lh)^2} (\lh\p\l) \lh_\a \right) Q W^\a  .
\label{p25}
\end{align}

Recalling (\ref{pure462}), it is
\begin{align}
b_{-1}^{(1)} U &= -\left( \frac1{8(\l\lh)^2} (r\g^{abc}\lh) ~ \left( N_{bc} ~ \l^\a \p_a A_\a \right) \right) \cr
&- \left( \frac1{16(\l\lh)^2} (r\g^{abc}\lh) ~ \left( \Pi_a ~ (\l\g_{bc})^\a A_\a \right) \right) \cr
&-\left(  \frac1{8(\l\lh)^3} (\l\g^{ab}r)  (\g_a\lh)^\a (\g_b\lh)^\b ~ \left( d_\a ~ \l^\g D_\b A_\g \right) \right) \cr
&+\p\left( \frac1{16(\l\lh)^2} (r\g^{abc}\lh) \right)  (\l\g_{bc})^\a \p_a A_\a  \cr
&-\p\left(\frac1{16(\l\lh)^3} (\l\g^{ab}r)  (\g_a\lh)^\a (\g_b\lh)^\b \right)  \l^\g D_\a D_\b A_\g .
\label{p26}
\end{align}
Considering the third line of this equation, it is equal to
\begin{align}
-&\left( \frac1{8(\l\lh)^3} (\l\g^{ab}r)  (\g_a\lh)^\a (\g_b\lh)^\b ~ Q \left( d_\a ~ A_\b \right) \right)  \cr
-&\left( \frac1{8(\l\lh)^3} (\l\g^{ab}r)  (\g_a\lh)^\a (\g_b\lh)^\b ~   \left( (\l\g^c)_\a \Pi_c ~ A_\b  \right) \right)  \cr
-&\left( \frac1{8(\l\lh)^3} (\l\g^{ab}r)  (\g_a\lh)^\a (\g_b\lh)^\b ~   \left( d_\a ~ (\l\g^c)_\a A_c \right) \right)  \cr
=-&\left( \frac1{8(\l\lh)^3} (\l\g^{ab}r)  (\g_a\lh)^\a (\g_b\lh)^\b ~ Q \left( d_\a ~ A_\b \right) \right)  \cr
-&\left( \frac1{8(\l\lh)^3} (\l\g^{bc}r)  (\lh\g_b\g^a\l) (\g_c\lh)^\a ~ \left( \Pi_a ~ A_\a \right) \right) \cr
-&\left( \frac1{8(\l\lh)^3} (\l\g^{bc}r) (\lh\g_c\g^a\l) (\g_b\lh)^\a ~ \left( d_\a ~ A_a\right) \right) \cr
+&\frac1{8(\l\lh)^3} (\l\g^{bc}r) (\lh\g_b\g^a\p\l) (\g_c\lh)^\a \p_a A_\a .
\label{p27}
\end{align}
The term with $\Pi$ in (\ref{p26}) and the term with $\Pi$ in (\ref{p27}) add to
\begin{align}
-\left( Q\left( \frac1{2(\l\lh)} (\g^a\lh)^\a \right) ~ \left( \Pi_a ~ A_\a \right) \right) ,
\label{p28}
\end{align}
where the identity (\ref{pure41}) was used. Consider now the first term in (\ref{p26}), after using (\ref{pure4}) it is equal to
\begin{align}
-&\left( \frac1{8(\l\lh)^2} (r\g^{abc}\lh) ~ Q \left( N_{bc} ~ A_a \right) \right) \cr
+ &\left( \frac1{8(\l\lh)^2} (r\g^{abc}\lh) ~ \left( \frac12 (\l\g_{bc} d) ~ A_a \right) \right) \cr
+&\left( \frac1{8(\l\lh)^2} (r\g^{abc}\lh) ~ \left( N_{bc} ~ (\l\g_a W) \right) \right) \cr
=-&\left( \frac1{8(\l\lh)^2} (r\g^{abc}\lh) ~ Q \left( N_{bc} ~ A_a \right) \right) \cr
+&\left( \frac1{16(\l\lh)^2} (r\g^{abc}\lh) (\l\g_{bc})^\a ~ \left( d_\a ~ A_a \right) \right) \cr
+&\frac1{16(\l\lh)^2} (r\g^{abc}\lh) (\p\l\g_{bc})^\a D_\a A_a \cr
+&\left( \frac1{8(\l\lh)^2} (r\g^{abc}\lh) ~ \left( N_{bc} ~ (\l\g_a W) \right) \right) .
\label{p29}
\end{align}
The term with $d$ here adds to the sixth line of (\ref{p27}) to produce
\begin{align}
\left( Q \left( \frac1{2(\l\lh)} (\g^a\lh)^\a \right) ~ \left( d_\a ~ A_a \right) \right) .
\label{p30}
\end{align}
Up to now we have
\begin{align}
b_{-1}^{(1)} U = &-\left( \frac1{8(\l\lh)^3} (\l\g^{ab}r)  (\g_a\lh)^\a (\g_b\lh)^\b ~ Q \left( d_\a ~ A_\b \right) \right)  \cr
&-\left( \frac1{8(\l\lh)^2} (r\g^{abc}\lh) ~ Q \left( N_{bc} ~ A_a \right) \right) \cr
&+ \left( Q \left( \frac1{2(\l\lh)} (\g^a\lh)^\a \right) ~\left( \left( d_\a ~ A_a \right)  - \left( \Pi_a ~ A_\a \right)  \right) \right) \cr
&+\left( \frac1{8(\l\lh)^2} (r\g^{abc}\lh) ~ \left( N_{bc} ~ (\l\g_a W) \right) \right) \cr
&+\frac1{8(\l\lh)^3} (\l\g^{bc}r) (\lh\g_b\g^a\p\l) (\g_c\lh)^\a \p_a A_\a \cr
&+ \frac1{16(\l\lh)^2} (r\g^{abc}\lh) \left(  \p\l\g_{abc} W \right)   \cr
&-\frac1{16(\l\lh)^2} (r\g^{abc}\lh) (\l\g_{bc})^\a \p(\p_a A_\a) \cr
&-\p\left(\frac1{16(\l\lh)^3} (\l\g^{ab}r)  (\g_a\lh)^\a (\g_b\lh)^\b \right)  \l^\g D_\a D_\b A_\g ,
\label{p31}
\end{align}
where total derivative terms are ignored. Consider the last term. After commuting the $D$'s and using (\ref{pure4}) this term is equal to
\begin{align}
&-\p\left(\frac1{16(\l\lh)^3} (\l\g^{ab}r)  (\g_a\lh)^\a (\g_b\lh)^\b \right) Q ( D_\a A_\b ) \cr
&+\frac1{8(\l\lh)^3} (\l\g^{bc}r) (\lh\g_c\g^a\p\l) (\g_b\lh)^\a \p_a A_\a \cr
&+\frac1{8(\l\lh)^3} (\l\g^{bc}r) (\lh\g_c\g^a\l) (\g_b\lh)^\a \p ( \p_a A_\a ) \cr
&+\frac1{16(\l\lh)^3} (\l\g^{bc}r) (\lh\g_c\g^a\p\l) (\lh\g_b\g_a W) \cr
&+\frac1{16(\l\lh)^3} (\l\g^{bc}r) (\lh\g_c\g^a\l) (\lh\g_b\g_a \p W) ,
\label{p310}
\end{align}
where a total derivative term has been ignored. Note that the term with $\p_a A_\a$ will cancel the fifth line of (\ref{p31}).

Consider now the combination
\begin{align}
\Delta \equiv \left( ( JW^\a + N^{ab}(\g_{ab}W)^\a ) ~~ Q \left( \frac1{4(\l\lh)} \lh_\a \right) \right) ,
\label{p32}
\end{align}
which will be related to the last term of (\ref{p31}). Applying $Q$, this expression becomes
\begin{align}
&\left( ( JW^\b + N^{ab}(\g_{ab}W)^\b ) ~ \left( \frac1{4(\l\lh)^2} r_\a \lh_\b ~~ \l^\a \right) \right) \cr
-&\left( ( JW^\a + N^{ab}(\g_{ab}W)^\a ) ~ \left( \frac1{4(\l\lh)^2} r_\a \lh_\b ~~ \l^\b \right) \right) .
\label{p33}
\end{align}
To use (\ref{p34}) the relevant OPE's  are
\begin{align}
( JW^\b + N^{ab}(\g_{ab}W)^\b ) (w) \frac1{4(\l\lh)^2} r_\a \lh_\b (z) &\to \frac1{(w-z)} \left( -\frac2{(\l\lh)^2} r_\a (\lh ~ W) \right) \cr
( JW^\a + N^{ab}(\g_{ab}W)^\a ) (w) \frac1{4(\l\lh)^2} r_\a \lh_\b (z) &\to \frac1{(w-z)} \left( -\frac2{(\l\lh)^3} (\l r) \lh_\b (\lh ~ W) \right) .
\label{p35}
\end{align}
Using these results, the expression (\ref{p32}) becomes, up to a total derivative term,
\begin{align}
\Delta =& \left( \frac1{4(\l\lh)^2} r_\a \lh_\b ~ \left( ( JW^{[\a} + N^{ab}(\g_{ab}W)^{[\a} ) ~ \l^{\b]} \right) \right) \cr
&+\frac2{(\l\lh)^2} (r\p\l) (\lh ~ W) - \frac2{(\l\lh)^3} (\l r) (\lh\p\l) (\lh ~ W) .
\label{p36}
\end{align}
Because
\begin{align}
 \left( ( JW^{[\a} + N^{ab}(\g_{ab}W)^{[\a} ) ~~ \l^{\b]} \right) = \frac1{48} \g_{abc}^{\a\b} \left( ( JW^\g + N^{cd}(\g_{cd}W)^\g ) ~~ \g^{abc}_{\g\r} \l^\r \right) ,
\label{p37}
\end{align}
one obtains
\begin{align}
\Delta =&  \left( \frac1{192(\l\lh)^2} (r\g_{abc}\lh) ~ \left ( ( JW^\a + N^{de}(\g_{de}W)^\a ) ~ \g^{abc}_{\a\b} \l^\b \right) \right) \cr
&+\frac2{(\l\lh)^2} (r\p\l) (\lh ~ W) - \frac2{(\l\lh)^3} (\l r) (\lh\p\l) (\lh ~ W) .
\label{p38}
\end{align}
but
\begin{align}
&\left(  ( JW^\a + N^{de}(\g_{de}W)^\a ) ~ \g^{abc}_{\a\b} \l^\b \right)  = - \left( J ~ (\l \g^{abc} W) \right) - \left( N_{de} ~ (\l\g^{abc}\g^{de} W) \right) \cr
&=24\left( N^{ab} ~ (\l \g^c W) \right) + 4 (\p\l\g^{abc} W) ,
\label{p39}
\end{align}
where the identity (\ref{p20}) was used.
Then, we have obtained
\begin{align}
&\left( ( JW^\a + N^{ab}(\g_{ab}W)^\a ) ~~ Q \left( \frac1{4(\l\lh)} \lh_\a \right) \right) =  \left( \frac1{8(\l\lh)^2} (r\g_{abc}\lh) ~ \left( N^{bc} ~ (\l\g^a W) \right) \right) \cr
&+\frac{1}{48(\l\lh)^2} (r\g_{abc}\lh) (\p\l \g^{abc} W) +\frac2{(\l\lh)^2} (r\p\l) (\lh ~ W) - \frac2{(\l\lh)^3} (\l r) (\lh\p\l) (\lh ~ W) .
\label{p40}
\end{align}
Using this result and (\ref{p310}) we obtain
\begin{align}
b_{-1}^{(1)} U = &-\left( \frac1{8(\l\lh)^3} (\l\g^{ab}r)  (\g_a\lh)^\a (\g_b\lh)^\b ~ Q \left( d_\a ~ A_\b \right) \right)  \cr
&-\left( \frac1{8(\l\lh)^2} (r\g^{abc}\lh) ~ Q \left( N_{bc} ~ A_a \right) \right) \cr
&+ \left( Q \left( \frac1{2(\l\lh)} (\g^a\lh)^\a \right) ~\left( \left( d_\a ~ A_a \right)  - \left( \Pi_a ~ A_\a \right)  \right) \right) \cr
&-\p\left(\frac1{16(\l\lh)^3} (\l\g^{ab}r)  (\g_a\lh)^\a (\g_b\lh)^\b \right) Q ( D_\a A_\b ) \cr
&+\left( ( JW^\a + N^{ab}(\g_{ab}W)^\a ) ~~ Q \left( \frac1{4(\l\lh)} \lh_\a \right) \right) \cr
&-Q \left( \frac1{2(\l\lh)} (\g^a\lh)^\a \right) \p ( \p_a A_\a ) - Q \left( \frac1{(\l\lh)} \lh_\a \right) \p W^\a  \cr
&+Q\left( \frac1{(\l\lh)^2} (\lh\p\l) \lh_\a \right) W^\a .
\label{p401}
\end{align}

Consider now (\ref{pure622}),
\begin{align}
b_{-1}^{(2)} U &= \left( \frac1{32(\l\lh)^4}  (\l\g^{ab}r) (r\g_{bcd}\lh) (\g_a\lh)^\a ~ \left( N^{cd} ~ \l^\b D_\a A_\b \right) \right)  \cr
&- \left( \frac1{64(\l\lh)^4}  (\l\g^{ab}r) (r\g_{bcd}\lh) (\g_a\lh)^\a ~ \left( d_\a ~ (\l\g^{cd})^\b A_\b \right) \right) \cr
&-\p\left( \frac1{64(\l\lh)^4}  (\l\g^{ab}r) (r\g_{bcd}\lh) (\g_a\lh)^\a \right)   (\l\g^{cd})^\b D_\a A_\b .
\label{p41}
\end{align}
Using (\ref{pure4}), the first line becomes
\begin{align}
&-\left( \frac1{32(\l\lh)^4}  (\l\g^{ab}r) (r\g_{bcd}\lh) (\g_a\lh)^\a ~ Q \left( N^{cd} ~ A_\a \right) \right) \cr
&+\left( \frac1{64(\l\lh)^4}  (\l\g^{ab}r) (r\g_{bcd}\lh) (\g_a\lh)^\b (\l\g^{cd})^\a ~ \left( d_\a ~ A_\b \right) \right) \cr
&+\left( \frac1{32(\l\lh)^4}  (\l\g^{ab}r) (r\g_{bcd}\lh) (\g_a\lh)^\a ~ \left( N^{cd} ~ (\l\g^e)_\a A_e \right) \right) \cr
&+\frac1{64(\l\lh)^4}  (\l\g^{ab}r) (r\g_{bcd}\lh) (\g_a\lh)^\a (\p\l\g^{cd})^\b D_\b A_\a ,
\label{p44}
\end{align}
the second line combines with the second line in (\ref{p41}) to produce
\begin{align}
\left( Q\left( \frac1{8(\l\lh)^3} (\l\g^{ab}r) (\g_a\lh)^\a (\g_b\lh)^\b \right) ~ \left( d_\a ~ A_\b \right) \right) .
\label{p45}
\end{align}
The last line here combines with the last term in (\ref{p41}) to give
\begin{align}
-Q \left( \frac1{16(\l\lh)^3} (\l\g^{ab}r) (\g_a\lh)^\a (\g_b\lh)^\b \right) \p ( D_\a A_\b) ,
\label{p450}
\end{align}
where the identity (\ref{pure41}) and the pure spinor conditions are used.

After using (\ref{p34}) and noting that
\begin{align}
&N^{cd}(w) \frac1{(\l\lh)^4} (\l\g^{ab}r) (r\g_{bcd}\lh)(\g_a\lh)^\a \to \mbox{regular} ,\cr
&N^{cd}(w) \frac1{(\l\lh)^4} (\l\g^{ab}r) (r\g_{bcd}\lh)(\g_a\lh)^\a (\l\g^e)_\a \to \mbox{regular} ,
\label{p46}
\end{align}
we obtain that the third term of (\ref{p44}) is equal to
\begin{align}
&\left( N^{cd} ~ \left( \frac1{32(\l\lh)^4}  (\l\g^{ab}r) (r\g_{bcd}\lh) (\g_a\lh)^\a ~~ (\l\g^e)_\a A_e \right) \right) \cr
&=\left( N^{cd} ~ \left( \frac1{32(\l\lh)^4}  (\l\g^{ab}r) (r\g_{bcd}\lh) (\g_a\lh)^\a(\l\g^e)_\a ~~  A_e \right) \right) \cr
&=\left(  \frac1{32(\l\lh)^4}  (\l\g^{ab}r) (r\g_{bcd}\lh) (\g_a\lh)^\a(\l\g^e)_\a ~ \left( N^{cd} ~ A_e \right) \right) \cr
&=\left(  Q \left( \frac1{8(\l\lh)^2} (r\g^{abc}\lh)  \right) ~  \left( N_{bc} ~ A_a \right) \right) .
\label{p47}
\end{align}
Then,
\begin{align}
b_{-1}^{(2)} U &=-\left( \frac1{32(\l\lh)^4}  (\l\g^{ab}r) (r\g_{bcd}\lh) (\g_a\lh)^\a ~ Q \left( N^{cd} ~ A_\a \right) \right) \cr
&+\left( Q\left( \frac1{8(\l\lh)^3} (\l\g^{ab}r) (\g_a\lh)^\a (\g_b\lh)^\b \right) ~ \left( d_\a ~ A_\b \right) \right) \cr
&-Q \left( \frac1{16(\l\lh)^3} (\l\g^{ab}r) (\g_a\lh)^\a (\g_b\lh)^\b \right) \p ( D_\a A_\b) \cr
&+\left(  Q \left( \frac1{8(\l\lh)^2} (r\g^{abc}\lh)  \right) ~  \left( N_{bc} ~ A_a \right) \right) .
\label{p48}
\end{align}

Finally, focusing on (\ref{pure8112}),
\begin{align}
b_{-1}^{(3)} U &= \left(  \frac1{256(\l\lh)^5} (\l\g^{ab}r) (r\g_{acd}\lh) (r\g_{bef}\lh) ~ \left( N^{cd} ~ \left( (\l\g^{ef})^\a A_\a \right) \right) \right) \cr
&-\p\left( \frac1{1024(\l\lh)^5} (\l\g^{ab}r) (r\g_{acd}\lh) (r\g_{bef}\lh)  \right) ~  (\l\g^{cd}\g^{ef})^\a A_\a \cr
&=-\left( Q\left( \frac1{32(\l\lh)^4}  (\l\g^{ab}r) (r\g_{bcd}\lh) (\g_a\lh)^\a \right) ~ \left( N^{cd} ~ A_\a \right) \right) .
\label{p49}
\end{align}

Therefore, adding (\ref{p25}), (\ref{p401}), (\ref{p48}) and (\ref{p49})  we finally obtain, up to total derivatives and $Q$ exact terms
\begin{align}
b_{-1} U = \p\t^\a A_\a + \Pi^a A_a + d_\a W^\a + \frac12 N^{ab} F_{ab} ,
\label{PF}
\end{align}
which is the expected result.

\section{Conclusion and prospects}
\label{conclu}

This work contains the complete calculation of the OPE between the
non-minimal $b$ ghost with the unintegrated vertex operator for the
first state in open superstring. The second order pole vanishes in
the Lorenz gauge and the first order pole gives the integrated vertex
operator. The computation is rather lengthily and is an important check
of formalism.

There are interesting questions regarding the composite $c$ ghost. In
the case of the bosonic string the unintegrated vertex operator for
the massless state in the Lorenz gauge can be written just as
$U=cV$. One could ask if such relation is also true for the pure
spinor formalism. Looking at the candidate $c$ ghost given in
\cite{Jusinskas:2013sha}
\begin{align}
  c=-\frac{r\lambda}{\partial\bar\lambda \lambda + r\partial\theta},
\end{align}
one can see that such relation cannot be true for the usual
unintegrated vertex operator. Any possible term depending on the
non-minimal fields in a physical $U$ is BRST exact. To prove the
relation $U=cV$ in the Lorenz gauge using the above $c$ ghost  one
likely must use a non-minimal composite operator that
trivializes the BRST cohomology, such as
\begin{align}
  \xi= \frac{\bar\lambda\theta}{\bar\lambda\lambda -r\theta}.
\end{align}
Allowing operators with this property in the formalism is the
equivalent of going from the small to the large Hilbert space in the
RNS string. If we allow non-covariant descriptions there are even more
possibilities. The paper \cite{Jusinskas:2015eza} has a very
interesting discussion on composite ghost fields and the relation
between different descriptions of physical states in the context of DDF
operators in the pure spinor string.

The main cause of the rather lengthy computation presented here is the
many terms present in $b$. As was shown in \cite{Berkovits:2010zz} for
a large class of supergravity backgrounds, including $AdS_5\times
S^5$, the expression for $b$ can be greatly simplified. Although
calculating the integrated vertex operator using the $b$ ghost would
require a background field expansion computation, it is possible that
using the $b$ ghost method is simpler than $QV=\partial U$ if one is
interested in some particular state.

\vskip 0.2in
{\bf Acknowledgements}~ We would like to thank Nathan Berkovits and
Renann Lipinski Jusinskas for useful comments and suggestions. We
would also like to thank Gabriel Chand\'ia for assistance with the
figure of this paper. This work  is partially financed by FONDECYT
grants 1200342 and 1201550.


{\small
\bibliographystyle{abe}
\bibliography{mybib}{}
}
\end{document}